# Comment on "A 'local observables' method for wave mechanics applied to atomic hydrogen," by Peter Bowman [Am. J. Phys. 76, 1120-1129 (2008)].

Hans C. Ohanian[a)]
*Department of Physics, University of Vermont, Burlington, Vermont 05405*

In a baffling paper[1] Bowman advocates that we abandon operators in quantum mechanics and calculate the angular momentum and magnetic moment of electron states directly from the electron currents in the wave fields, treating these as though they were classical quantities. He performs such a calculation for the states of the hydrogen atom and makes the bizarre claim that the total angular momentum of the ground state is $\hbar$, rather than $\hbar/2$.

It is obvious that this result must be wrong. The ground state wavefunction is spherically symmetric in its dependence on the spatial coordinates, so the angular momentum has no orbital contribution and must be entirely attributed to the electron's spin. If the angular momentum of the ground state were $\hbar$, then the spin of the electron would have to be $\hbar$;[2] that is, the electron would have to be what we conventionally call a particle of spin 1. This spin value for the electron would lead to disastrous mathematical and physical consequences. It would mean that Bowman's calculation is self-contradictory, because he uses the Dirac equation appropriate to spin 1/2 and not the Proca equation[3] required for a massive particle of spin 1. The spectral lines of hydrogen and alkali atoms would display fine-structure triplets, instead of the observed doublets. And the gyromagnetic ratio of the electron would be 1, instead of the observed value of 2.00…, which is found from measurements of the electron magnetic resonance frequency.

If Bowman's result is wrong, where is the mistake in his calculation? It turns out that the mistake is elementary. He attributes the angular momentum to a circulatory flow of mass, obtained by multiplying the Pauli electron current density $\mathbf{k}_{pauli}$ by the mass:

$$m_e \mathbf{k}_{pauli} = \frac{\hbar}{2}[2\,\text{Im}(\psi^\dagger \nabla \psi) + \nabla \times (\psi^\dagger \sigma \psi)] \tag{1a}$$

$$= \frac{\hbar}{2i}[\psi^\dagger \nabla \psi - (\nabla \psi^\dagger)\psi] + \frac{\hbar}{2}\nabla \times (\psi^\dagger \sigma \psi) \tag{1b}$$

The Pauli electron current density is the non-relativistic limit of the Dirac electron current density, and the use of a non-relativistic approximation in hydrogen-like atoms (in which the electron speeds are as large as $c/137$, and even larger for atoms of high $Z$) is questionable. But Bowman's mistake is not this non-relativistic approximation, but a confusion over the source of angular momentum. The electron current density (or some current density proportional to it) is not the source of angular momentum. Instead, the source of angular momentum is the momentum density, which is given by the $T^{0k}$ components of the energy-momentum tensor (what in electromagnetic theory we would call the Poynting vector). *This momentum density is not equal to the electron current density multiplied by the mass.*

In a paper published in 1986 (which Bowman references but evidently chose to ignore),[4] I performed a calculation similar to Bowman's, with the goal of providing an intuitive interpretation of the spin. In this calculation I used the correct (and relativistically exact) momentum density **G** for the Dirac field, extracted from the symmetrized energy-momentum tensor:[5]

$$\mathbf{G} = \frac{\hbar}{2i}[\psi^\dagger \nabla \psi - (\nabla \psi^\dagger)\psi] + \frac{\hbar}{4}\nabla \times (\psi^\dagger \sigma \psi) \tag{2}$$

The magnetic moment is the volume integral of $-\frac{1}{2}e\mathbf{x}\times\mathbf{k}_{pauli}$, whereas the angular momentum is the volume integral of $\mathbf{x}\times\mathbf{G}$. Comparison of Eqs. (1b) and (2) shows that the factor multiplying the spin term in Eq. (1b) is twice as large as in Eq. (2). Because of this extra factor of two, the electron spin contributes twice as much to the magnetic moment as it contributes to the angular momentum. This difference accounts for the non-classical gyromagnetic ratio associated with the electron spin. Hence Bowman's calculation of the magnetic moment of the ground state of the hydrogen atom is correct, but his calculation of the angular momentum is off by a factor of two. There are similar mistakes in his calculations for all the other states.

Bowman proposes that a direct measurement of the angular momentum of the ground state of hydrogen could serve as an *experimentum crucis* that discriminates between his "local observables" scheme and standard quantum mechanics. My correction of his calculation shows that such a measurement would do no such thing — the angular momentum of the ground state is $\hbar/2$, no matter which way we calculate it. Although it is instructive to calculate physical quantities by integration over the wave fields, these calculations (when done correctly) will always agree with the results obtained by the operator methods of standard quantum mechanics. Thus, there is no need for a reformulation of quantum mechanics.

**Acknowledgment**
I would like to acknowledge cordial and helpful correspondence from Peter Bowman about this comment.

---

[a)] Electronic mail: hohanian@uvm.edu

[1] P. Bowman, "A 'local observables' method for wave mechanics applied to atomic hydrogen," Am. J. Phys. **76**, 1120-1129 (2008).

[2] This result can be easily confirmed by applying Bowman's calculation to a free-electron wavefunction.

[3] G. Wentzel, *Quantum Theory of Fields* (Interscience Publishers, New York, 1949), p. 75.

[4] H. C. Ohanian, "What is spin?" Am. J. Phys. **54**, 500-505 (1986).

[5] The symmetrized energy-momentum tensor for the Dirac field is given by G. Wentzel, op. cit., p. 170. Wentzel uses a slightly different notation, with an extra factor of $i\beta$ included in $\psi^\dagger$. The tensor given by Wentzel can be rewritten in the form of Eq. (2) (the manipulations in Ref. 2 used the Dirac equation for a free electron, but it is easy to check that Eq. (2) remains valid when the Dirac equation includes electromagnetic fields). Symmetrization of the energy-momentum tensor was first proposed by F. J. Belinfante, "On the spin angular momentum of mesons," Physica **6**, 887-898 (1939). If instead of Belinfante's symmetric energy-momentum tensor, we adopt an asymmetric tensor such that the spin cannot be attributed to the moment of the momentum density, then an extra spin density must be added "by hand" to account for all or part of the electron spin. In essence, this spin density means that the electron is regarded as a spinning pointlike entity with an "internal" angular momentum (somewhat à là Goudsmit and Uhlenbeck). Such a description of the spin is strongly favored by some physicists who wish to use the spin density as the source for a torsion of spacetime.